\documentclass{appolb}
\usepackage{epsfig}
\usepackage{amsmath}
\newcommand{\ext}{\operatorname{ext}}
\begin{document}
\title{Spin-Gravity Coupling}
\author{Bahram Mashhoon
\address{Department of Physics and Astronomy\\University of Missouri-Columbia\\Columbia, Missouri 65211, USA}
}
\maketitle
\begin{abstract}
Mathisson's spin-gravity coupling and its Larmor-equivalent interaction, namely, the spin-rotation coupling are discussed. The study of the latter leads to a critical examination of the basic role of locality in relativistic physics. The nonlocal theory of accelerated systems is outlined and some of its implications are described.
\end{abstract}
  
\section{Introduction}

The proper theory of the motion of a spinning mass in a gravitational field is due to Mathisson~\cite{1}, Papapetrou~\cite{2}, and Dixon~\cite{3}. A main aspect of this theory, which already appears in the work of Mathisson~\cite{1}, is the existence of a spin-curvature force
\begin{equation}\label{eq:1} F^\alpha =-\frac{c}{2}R^\alpha_{\;\;\beta \mu \nu} u^\beta S^{\mu\nu}.\end{equation}
Here $u^\mu$ is the unit four-velocity vector of the spinning mass; that is, $u^\mu =dx^\mu /d\tau$, where $x^\mu =(ct,x,y,z)$ and $\tau /c$ is the proper time. The signature of the metric is $+2$ throughout this paper.

In the linear approximation of general relativity, with the spinning mass held at rest in the stationary exterior field of a rotating central source and keeping only first-order terms in spin, $F^\alpha=(0,\mathbf{F})$, where \cite{4}
\begin{equation}\label{eq:2} \mathbf{F}=-\mathbf{\nabla} (\mathbf{S}\cdot \mathbf{\Omega}_P).\end{equation}
Here $\mathbf{\Omega}_P$ is the precession frequency of a test gyroscope held at rest outside the central source of angular momentum $\mathbf{J}$; far from the source,
\begin{equation}\label{eq:3}\mathbf{\Omega}_P =\frac{G}{c^2r^5}[3(\mathbf{J}\cdot \mathbf{r})\mathbf{r}-\mathbf{J}r^2],\end{equation}
so that $c\mathbf{\Omega}_P=\mathbf{B}_g$ is the familiar dipolar gravitomagnetic field of the source.

It follows from Eq.~\eqref{eq:2} that one can define the Hamiltonian for the spin-gravity coupling as
\begin{equation}\label{eq:4}\mathcal{H}=\mathbf{S}\cdot \mathbf{\Omega}_P.\end{equation}
This Mathisson Hamiltonian is a direct analogue of $-\mathbf{\mu}\cdot \mathbf{B}$ coupling in electrodynamics~\cite{5}. Imagine now the test gyroscope that is held at rest but precesses with frequency $\mathbf{\Omega}_P$ as before. If the gravitational interaction is turned off, the gyro keeps its direction fixed with respect to the background global inertial frame by the principle of inertia. The former precessional motion is recovered, however, from the viewpoint of a local observer that is at rest in a frame of reference rotating with frequency $\mathbf{\Omega}=-\mathbf{\Omega}_P$. This is an instance of the gravitational Larmor theorem~\cite{5}, which follows from Einstein's principle of equivalence. To this latter motion in the rotating frame, one can associate a new Hamiltonian $\mathcal{H}'$, which can be obtained from $\mathcal{H}$ by replacing $\mathbf{\Omega}_P$ with $-\mathbf{\Omega}$. Thus the Hamiltonian due to the coupling of spin with rotation is given by
\begin{equation}\label{eq:5}\mathcal{H}'=-\mathbf{S}\cdot \mathbf{\Omega}.\end{equation}
The classical couplings \eqref{eq:4} and \eqref{eq:5} are expected to be valid for intrinsic spin as well. This is mainly based on the study of relativistic wave equations in gravitational fields and accelerated frames of reference, see~\cite{6} for some examples; a more complete discussion as well as list of references is given in \cite{7}.

It follows from the inertia of intrinsic spin that to every spin Hamiltonian in a laboratory fixed on the Earth, one must add
\begin{equation}\label{eq:6} \delta \mathcal{H}\approx -\mathbf{S}\cdot \mathbf{\Omega}_\oplus +\mathbf{S}\cdot \mathbf{\Omega}_{P\oplus}.\end{equation}
For a spin-$\frac{1}{2}$ particle, the spin-rotation part of Eq.~\eqref{eq:6} implies that the maximum energy difference between spin-up and spin-down states is $\hbar \Omega_\oplus \approx 10^{-19}$eV. As pointed out in~\cite{8}, the experimental results of \cite{9} constitute an indirect measurement of this coupling. Further evidence in this direction, based on an analysis of the muon $g-2$ experiment, is discussed in \cite{10}; for other observational aspects of spin-rotation coupling see \cite{7}. Moreover, the corresponding energy difference for the spin-gravity term in Eq.~\eqref{eq:6} is $\hbar \Omega_{P\oplus}\approx 10^{-29}$eV. As discussed in \cite{7}, even in a space-borne laboratory in orbit around Jupiter, this Mathisson coupling would still be too small to be measurable at present by several orders of magnitude. An interesting recent discussion of the theoretical as well as observational aspects of spin-gravity coupling is contained in \cite{11}.

Finally, a fundamental aspect of the Mathisson coupling should be noted here: For a classical gyro, its spin is proportional to its mass and the gravitational force \eqref{eq:2} is then proportional to the mass of the gyro, as it should be; however, for a spin-$\frac{1}{2}$ particle, the magnitude of spin is $\hbar /2$ and the corresponding gravitational Stern-Gerlach force \eqref{eq:2} violates the universality of free fall. Thus the weight of a neutron with spin up is generally different from the weight of the same neutron with spin down; however, this effect is too small to be measurable in the foreseeable future~\cite{7}. Nevertheless, this observation indicates that the simple coupling \eqref{eq:4} for intrinsic spin as well as its Larmor-equivalent \eqref{eq:5} could have consequences that are of basic significance for relativity theory and gravitation. This important point constitutes the main theme of this paper and will be illustrated in subsequent sections. In practice, it is indeed much simpler to work with \eqref{eq:5} than with \eqref{eq:4}; therefore we concentrate on the photon spin-rotation coupling in the rest of this paper.

\section{Photon helicity-rotation coupling}\label{s:2}

Consider a thought experiment in which an observer rotates uniformly with frequency $\Omega$ about the direction of propagation of an incident plane monochromatic electromagnetic wave of frequency $\omega$. The object of the experiment is to measure $\omega '$, the wave frequency according to the rotating observer. Specifically, we assume that the wave propagates along the $z$ direction and the observer follows a circle of radius $r$ about the origin of spatial coordinates in the $(x,y)$ plane. The natural orthonormal tetrad frame associated with the observer is given by
\begin{align}\label{eq:7} \lambda^\mu _{\;\; (0)}&= \gamma (1,-\beta \sin \varphi ,\beta \cos \varphi ,0),\\
\label{eq:8} \lambda^\mu _{\;\;(1)} &= (0,\cos \varphi,\sin \varphi,0),\\
\label{eq:9} \lambda^\mu_{\;\;(2)} &= \gamma (\beta ,-\sin \varphi ,\cos \varphi ,0),\\
\label{eq:10} \lambda^\mu_{\;\;(3)}&=(0,0,0,1).\end{align}
Here $\varphi =\Omega t=\gamma \Omega \tau /c$, $\beta =r\Omega /c$, and $\gamma =(1-\beta ^2)^{-1/2}$. The observer's local temporal axis is along its four-velocity $\lambda^\mu _{\;\;(0)}$ and its spatial frame $\lambda^\mu_{\;\;(i)}$, $i=1,2,3$, is such that its axes point along the radial, tangential, and $z$ directions, respectively.

According to the standard Doppler effect, the frequency of the wave measured by the observer is $\omega '_D=-k_\mu \lambda ^\mu_{\;\;(0)} =\gamma \omega$, where the Lorentz factor accounts for time dilation. In this general approach, the rotating observer is assumed to be pointwise inertial and hence at rest in a comoving inertial frame (``hypothesis of locality") and the Doppler effect follows from the invariance of the phase of the wave under Lorentz transformations between the global background inertial frame and the instantaneous inertial frames of the observer.

There is, however, another way to measure frequency based on the fact that at least a few periods of the wave must be registered before the observer can determine $\omega '$. To this end, we suppose that the observer can make pointwise determinations of the incident field. The result can be expressed in terms of instantaneous Lorentz transformations or equivalently as
\begin{equation}\label{eq:11} F_{(\alpha )(\beta)}(\tau ) =F_{\mu\nu}\lambda^\mu_{\;\;(\alpha )} \lambda^\nu_{\;\;(\beta)} .\end{equation}
This quantity, upon Fourier analysis, yields \cite{12}
\begin{equation}\label{eq:12} \omega '=\gamma (\omega \mp \Omega).\end{equation}
The upper (lower) sign refers to an incident positive (negative) helicity wave. For the photon energy, we find that
\begin{equation}\label{eq:13} E'=\gamma (E\mp \hbar\Omega ),\end{equation}
where $\pm \hbar$ is the photon helicity. Thus Eqs. \eqref{eq:12} and \eqref{eq:13} contain, in addition to the transverse Doppler effect, the influence of the spin-rotation coupling. Eq.~\eqref{eq:12} can be written as $\omega '=\omega'_D (1\mp \Omega /\omega)$, where $\Omega /\omega$ is the ratio of the reduced wavelength of the radiation $\lambda /(2\pi)$ to the acceleration length $\mathcal{L}$ of the observer, $\mathcal{L}=c/\Omega$. The Doppler effect is recovered when this ratio vanishes in the JWKB limit.

For oblique incidence, the analogue of Eq.~\eqref{eq:13} is
\begin{equation}\label{eq:14} E'=\gamma (E-\hbar M\Omega),\end{equation}
where $\hbar M$ is the total angular momentum of the radiation along the axis of rotation. Thus $\omega '=\gamma (\omega -M\Omega)$, where $M=0,\pm 1,\pm 2,\dots$, for a scalar or a vector field, while $M\mp \frac{1}{2}=0,\pm 1,\pm 2,\dots$, for a Dirac field. In the JWKB approximation, Eq.~\eqref{eq:14} may be expressed as $E'=\gamma(E-\mathbf{J} \cdot \mathbf{\Omega})$; hence, $E'=\gamma (E-\mathbf{v}\cdot\mathbf{p})-\gamma \mathbf{S}\cdot\mathbf{\Omega}$, where $\mathbf{J}=\mathbf{r}\times \mathbf{p}+\mathbf{S}$ and $\mathbf{v}=\mathbf{\Omega}\times\mathbf{r}$. It is important to note that $\omega '$ vanishes for $\omega =M\Omega$, while $\omega '$ can be negative for $\omega <M\Omega$. The former circumstance poses a basic difficulty, while the latter is a consequence of the absolute character of accelerated motion~\cite{12}.

It is useful to provide an intuitive explanation for the appearance of the spin-rotation term in Eq.~\eqref{eq:12}. In an incident positive (negative) helicity wave, the electric and magnetic fields rotate with frequency $\omega$ in the positive (negative) sense about the direction of propagation of the wave. The observer rotates about this direction with frequency $\Omega$; therefore, relative to the observer, the electric and magnetic fields of the incident wave rotate with frequency $\omega -\Omega\; (\omega +\Omega)$ in the positive (negative) helicity case. While the relative circular motion accounts for the subtraction (addition) of frequencies, the Lorentz factor in Eq.~\eqref{eq:12} takes care of time dilation. This factor is unity for the rotating observer at $r=0$, hence $\omega '=\omega \mp \Omega$ in this case; the fact that only the Lorentz factor distinguishes rotating observers at different radii in Eq.~\eqref{eq:12} follows intuitively from the circumstance that each such observer is locally equivalent to the one at $r=0$, since each is locally a center of rotation of frequency $\Omega$.

The existence of spin-rotation coupling in Eq.~\eqref{eq:12} can be observationally demonstrated by various means including the GPS, where it accounts for the phenomenon of phase wrap-up. That is, for $\gamma \ll 1$ and $\Omega \ll\omega$, $\omega '\approx\omega \mp \Omega$ has been verified with $\omega /(2\pi)\sim 1\text{ GHz}$ and $\Omega /(2\pi)\sim 8\text{ Hz}$ \cite{13}. Further observational aspects of Eq.~\eqref{eq:12} are discussed in \cite{14}.

The exact result $\omega '=\gamma (\omega -\Omega)$ for incident positive-helicity radiation has a fundamental consequence that must now be addressed. This relation implies that $\omega '=0$ for $\omega =\Omega$. The incident radiation stands completely still with respect to all observers that uniformly rotate with frequency $\omega$ about the direction of propagation of the wave. That by a mere rotation an observer can stand still with an electromagnetic wave is analogous to the pre-relativistic formula for the Doppler effect where an observer moving with speed $c$ along a beam of light would see an electromagnetic field that is spatially oscillatory but at rest. This paradoxical circumstance played a role in Einstein's path to relativity theory (see p. 53 of \cite{15}, which contains Einstein's autobiographical notes). The origin of this defect in Eq.~\eqref{eq:12} must be sought in Eq.~\eqref{eq:11}, namely the assumption that the field measured by the rotating observer is pointwise the same as that measured by the momentarily comoving inertial observer (``hypothesis of locality"); a brief critique of this notion of locality is contained in the next section. The other nonlocal assumption, involving the Fourier analysis of the measured field, is reasonable, since a number of periods of the wave must be received by the accelerated observer before $\omega '$ could be adequately measured.

\section{Hypothesis of locality}\label{s:3}

According to the standard theory of relativity, Lorentz invariance is extended to accelerated observers in Minkowski spacetime via the hypothesis of locality, namely, the assumption that an accelerated observer, at each instant along its worldline, is momentarily equivalent to an otherwise identical hypothetical comoving inertial observer. For time determination, this assumption reduces to the clock hypothesis. Thus an accelerated observer is pointwise inertial and this supposition provides operational significance for Einstein's principle of equivalence \cite{16}.

Regarding the source of this important postulate of relativity theory, it must be noted that Lorentz introduced it as an approximation in his discussion of the Lorentz-Fitzgerald contraction of electrons in curvilinear motion (see section 183 of \cite{17}). Einstein mentioned it in his discussion of accelerated systems (see p. 60 of \cite{18}). Weyl likened it to the assumption of adiabaticity in thermodynamics (see pp. 176-177 of \cite{19}).

The locality assumption originates from Newtonian mechanics, where the state of a particle is determined by its position and velocity. The accelerated observer shares the same state with the comoving inertial observer; hence, locality  is exact and no new physical assumption is needed if all physical phenomena could be reduced to pointlike coincidences of classical particles and null rays. However, when wave phenomena are taken into consideration, the locality hypothesis would be approximately valid whenever $\lambda \ll \mathcal{L}$. Here $\lambda$ is the characteristic wavelength of the phenomena under observation and $\mathcal{L}$, the acceleration length, is the characteristic length scale for the variation of the state of the observer. In practice, deviations from locality are expected to be of order $\lambda /\mathcal{L}$ and are generally very small, since $\mathcal{L}$ is quite long; for instance, $c^2/g_\oplus \approx 1$ lyr and $c/\Omega_\oplus \approx 28$ AU for an observer in a laboratory fixed on the Earth. The consistency of these ideas can be illustrated by two examples of general interest.

Imagine a classical charged particle of mass $m$ and charge $q$ that is subject to an external force $\mathbf{F}_{\ext}$. The accelerated charge radiates electromagnetic radiation with characteristic wavelength $\lambda \sim \mathcal{L}$. The hypothesis of locality is thus violated since $\lambda /\mathcal{L}\sim 1$. This means that the state of the charged particle cannot be given at each instant by its position and velocity alone. This is consistent with the equation of motion of the particle, which reduces to the Abraham-Lorentz equation
\begin{equation}\label{eq:15} m\frac{d\mathbf{v}}{dt}-\frac{2}{3} \frac{q^2}{c^3} \frac{d^2\mathbf{v}}{dt^2}+\dots =\mathbf{F}_{\ext}\end{equation}
in the nonrelativistic  approximation.

Consider next muon decay in a storage ring \cite{20}. This experiment has verified with good accuracy relativistic time dilation $\tau _\mu=\gamma\tau _\mu^0$, where $\tau _\mu^{0}$ is the lifetime of the muon at rest. To mimic the circular acceleration of a muon in a storage ring and take the quantum nature of this particle into account, one can suppose that the muon decays from a high-energy Landau level in a constant magnetic field. Based on the detailed calculation reported in \cite{21},
\begin{equation}\label{eq:16} \tau _\mu \approx \gamma \tau _\mu ^{ 0} \left[ 1+\frac{2}{3} \left( \frac{\lambda _C}{\mathcal{L}}\right)^2\right],\end{equation}
where $\lambda_C$ is the Compton wavelength of the muon and $\mathcal{L} =c^2/a$, where $a\sim 10^{18}g_\oplus$ is the effective acceleration of the muon. The correction to the standard formula in Eq. \eqref{eq:16} is very small $(\sim 10^{-25})$, but nonzero.

\section{Nonlocality}\label{s:4}

To go beyond the hypothesis of locality, let us return to Eq. \eqref{eq:11} and consider its generalization. Let $\mathcal{F}_{(\alpha )(\beta )} (\tau )$ be the field that is actually measured by the accelerated observer. Here $\tau$ is measured by the background inertial observers using $d\tau =cdt/\gamma$. The most general linear relationship between $\mathcal{F}_{(\alpha )(\beta)} (\tau )$ and the field measured by the infinite sequence of comoving inertial observers, given by Eq. \eqref{eq:11}, that preserves causality is given by \cite{22}
\begin{equation}\label{eq:17} \mathcal{F}_{(\alpha )(\beta)} (\tau )=F_{(\alpha )(\beta)} (\tau )+\int^\tau_{\tau_0} K_{(\alpha )(\beta)}^{\;\;\;\;\;\;\;\;\;(\gamma )(\delta)} (\tau ,\tau ')F_{(\gamma )(\delta )}(\tau ') d\tau '.\end{equation}
Here $\tau_0$ is the instant at which the acceleration is turned on and the kernel $K$ is such that it vanishes in the absence of acceleration. The integral in Eq.~\eqref{eq:17} has the form of an average over the past worldline of the accelerated observer; moreover, it is expected to vanish in the JWKB limit $(\lambda /\mathcal{L}\to 0)$. It is a consequence of the Volterra-Tricomi theorem~\cite{23} that under reasonable physical conditions the relationship between $\mathcal{F}_{(\alpha )(\beta)}$ and $F_{(\alpha )(\beta)}$ is unique.

How should the kernel be determined? This involves various complications~\cite{22}, but a key idea is that the kernel should be so chosen as to prevent the circumstance encountered in section~\ref{s:2}. That is, we introduce the fundamental postulate that a basic radiation field can never stand completely still with respect to an arbitrary observer. A detailed treatment of the nonlocal theory of accelerated systems is contained in~\cite{24} and the references cited therein. This theory is in agreement with available observational data; moreover, it forbids the existence of a fundamental scalar (or pseudoscalar) field.

What are the implications of nonlocality for the photon helicity-rotation coupling in the thought experiment of section~\ref{s:2}? There are basically two aspects of the problem that are altered by nonlocality:

(i) As determined by the rotating observer, for $\omega >\Omega$ the amplitude of the positive-helicity incident wave is enhanced, while the amplitude of the negative-helicity wave is diminished.

(ii) For $\omega =\Omega$, the field is not static in the positive helicity case; instead, it varies like $t$ as in the case of resonance.

It is important to verify these purely nonlocal effects experimentally. The task here is complicated by the fact that the behavior of rotating measuring devices must be known. An interesting discussion of such issues of principle is contained in \cite{25}. We therefore turn to a different approach based on the correspondence principle in nonrelativistic quantum mechanics. The study of electrons in rotational motion within the framework of quantum theory could shed light on the question of the correct classical theory of accelerated systems.

In connection with (i), the cross section $\sigma$ for the photoionization of hydrogen atom has been studied with the electron in a circular state with respect to the incident radiation that would correspond to the motion of the observer in section~\ref{s:2}. A detailed investigation reveals that $\sigma_+>\sigma_-$, where $\sigma _+ (\sigma_-)$ is the cross section in the case that the electron rotates in the same (opposite) sense as the helicity of the incident radiation~\cite{26}.

The situation in (ii) can be mimicked by the transition of an electron in a circular ``orbit" about a uniform magnetic field to the next energy state as a result of absorption of a photon of frequency $\Omega_c$ and definite helicity that is incident along the direction of the magnetic field. Here $\Omega_c$ is the electron cyclotron frequency. Let $P$ be the probability of transition to the next energy state. A detailed investigation reveals that in the correspondence regime, $P_+\propto t^2$, while $P_-=0$, corresponding to the positive and negative helicity cases, respectively~\cite{26}.

It appears from these studies that the nonlocal theory is in better agreement with quantum theory than the standard theory of relativity that is based on the hypothesis of locality~\cite{26}.

\section{Discussion}\label{s:5}

Mathisson's spin-gravity Hamiltonian leads, via the gravitational Larmor theorem, to the spin-rotation Hamiltonian. For the photon, helicity-rotation coupling has the consequence that a rotating observer can in principle be comoving with an electromagnetic wave such that the wave is oscillatory in space but stands completely still with respect to the observer. The source of this difficulty is the hypothesis of locality that is the basis for the extension of Lorentz invariance to accelerated observers and the subsequent transition to general relativity. The nonlocal theory of accelerated systems is briefly described; in this theory, instead of the locality assumption, where a curved worldline is in effect replaced at each instant by the straight tangent worldline, one considers in addition an average over the past worldline of the observer. The consequences of this nonlocal special relativity are briefly described. The nonlocal theory is in agreement with available observational data. It remains to extend this theory to a nonlocal theory of gravitation.

\section*{Acknowledgement}

I am grateful to Andrzej Trautman for his kind invitation to present
this work and warm hospitality at the Mathisson Conference (17-20 October
2007, Warsaw, Poland).

\end{document}